\documentclass[twocolumn,showpacs,preprintnumbers,amsmath,amssymb,superscriptaddress]{revtex4}

\usepackage{graphicx}% Include figure files
\usepackage{epstopdf}% Include figure files
\usepackage{dcolumn}% Align table columns on decimal point
\usepackage{bm}% bold math

\begin{document}
\title{Extrinsic Proton NMR Studies of Mg(OH)$_2$ and Ca(OH)$_2$}

\author{Yutaka Itoh}
 \affiliation{Department of Physics, Graduate School of Science, Kyoto Sangyo 
University, Kamigamo-Motoyama, Kika-ku, Kyoto 603-8555, Japan}
\author{Masahiko Isobe}
 \affiliation{Max-Planck-Institut f\"{u}r Festk\"{o}rperforschung, Heisenbergstrasse 1, D-70569 Stuttgart, Germany}

\begin{abstract}
We studied narrow $^1$H NMR spectra of Mg(OH)$_2$ and Ca(OH)$_2$ powders at 100$-$355 K and 42$-$59 MHz
using pulsed NMR techniques. 
The Fourier transformed NMR spectra of the proton free-induction signals show the superposition of broad and narrow components, which can be assigned to immobile protons and extrinsic mobile protons, respectively.  
We found that a narrow spectrum develops on heating above about $T_c$ = 260 K and widens above a Larmor frequency of about $\nu_c$ = 50 MHz for Mg(OH)$_2$.  
The temperature-induced NMR spectrum and the characteristic frequency $\nu_c$ of 50 MHz are the noteworthy features of the nuclear spin fluctuation spectra of the extrinsic protons.  
\end{abstract}
 
 \date{\today}
%%%

\pacs{correlated protons, NMR}

\maketitle
  
\section{Introduction}  
The conventional motional narrowing effects on NMR spectra are the rapid decrease in NMR linewidths on heating above specific temperatures due to molecular motion and atomic diffusion~\cite{Abragam}.  
Ionic conduction facilitated by lattice defects also causes the narrowing of the NMR spectra~\cite{Li0}.  
The development of a narrow NMR spectrum superposed with a broad NMR spectrum has been observed for some hydrates~\cite{Corey1,Corey2,Corey3,Corey4}, hydrogen-bonded compounds~\cite{TDA,KDP}, and nanocrystalline Li compounds~\cite{Li1,Li2,Li3}. 
The broad component is assigned to immobile ions with dipole-dipole couplings on a rigid lattice,
while the narrow component is assigned to mobile ions decoupled from the matrix of each material. 
Recent pulsed NMR studies of Mg(OH)$_2$ powder also showed the coexistence of narrow and broad $^1$H NMR spectra in a wide temperature range~\cite{Itoh}.  

Single-crystal $^1$H NMR studies of Mg(OH)$_2$ at room temperature showed wide NMR spectra and angular dependences with respect to the crystal axes~\cite{oNMR,sNMRCa}, while powder NMR studies showed narrow NMR spectra~\cite{pNMR,MAS}.
Both single-crystal and powder $^1$H NMR studies of Ca(OH)$_2$ showed wide NMR spectra~\cite{pNMRCa}.  
We revealed the superposition of broad and narrow NMR spectra for Mg(OH)$_2$, which    
were assigned to dipole-coupled protons on a rigid lattice and to extrinsic protons, respectively~\cite{Itoh}.   
 
\begin{figure}[b]
 \begin{center}
 \includegraphics[width=0.90\linewidth]{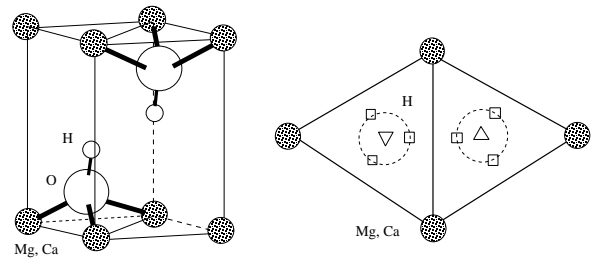}
 \end{center}
 \caption{\label{f1}
Crystal structure of divalent hydroxides M(OH)$_2$ (M = Mg and Ca).  
The right illustration is a top view of M(OH)$_2$. 
Open squares are the XGT sites with 1/3 occupation for protons.   
}
 \end{figure}
 
The crystal structure of the divalent hydroxides M(OH)$_2$ (M = Mg, Ca) is characterized by a layered structure and OH groups (trigonal, $P\overline{3}m1$). 
Figure~\ref{f1} shows a unit cell of the crystal lattice. 
The lattice spacing and unit cell volume of Ca(OH)$_2$ are larger than those of Mg(OH)$_2$~\cite{ND2}. 
According to neutron diffraction studies~\cite{NDCa,ND2}, the hydrogen site symmetry with a threefold axis is broken and
the space group of the crystal structure is $P\overline{3}$.
The location of the hydrogen atoms is assigned to the XGT sites with 1/3 occupation in Fig.~\ref{f1}. 

Mg(OH)$_2$ and Ca(OH)$_2$ are proton conductors above room temperature~\cite{Freund}. 
The thermal activation energy of proton transport was estimated to be 2.0 eV for Mg(OH)$_2$ and 2.2 eV for Ca(OH)$_2$~\cite{Freund}. 
Excess protons and proton vacancies are considered to be mobile carriers in the conductivity.
The extrinsic protons associated with the narrow NMR spectrum may be mobile protons facilitated by lattice defects.   
 
In this paper, we studied the narrow component of the proton NMR spectrum for commercially available powder samples of Mg(OH)$_{2}$ and Ca(OH)$_{2}$ by pulsed Fourier transformed (FT) $^{1}$H NMR techniques with variable temperature and Larmor frequency $\nu_\mathrm{L}$.
We found unconventional strong temperature and frequency dependences of the narrow component.  
The proton spin fluctuation spectrum with a characteristic temperature $T_c$ of 260 K and a characteristic frequency $\nu_c$ of 50 MHz is different from that obtained from a conventional Lorentzian model. 
To our knowledge, our discovery of the characteristic frequency $\nu_c$ of 50 MHz is unprecedented.  
  
\section{Experimental Procedure}

The present NMR experiments were carried out for commercially available powder samples of Mg(OH)$_2$ (99.9$\%$ purity, Wakenyaku)
and Ca(OH)$_2$ (99.995$\%$ purity, Sigma-Aldrich).
The samples were annealed at 120 $^\circ$C for 18 h in air 
and were confirmed to have a single phase from their powder X-ray diffraction patterns. 
The samples were treated in a glove box with N$_2$ gas (a dry atmosphere)
and were sealed in glass tubes with N$_2$ gas for NMR experiments.  
A phase-coherent-type pulsed spectrometer was used for the $^{1}$H proton NMR (nuclear spin $I$ = 1/2) experiments from a Larmor frequency $\nu_\mathrm{L}$ of 42.5772 MHz in an external magnetic field of 1.0 T to 59.6 MHz in a magnetic field of 1.4 T.    
The NMR frequency spectra were obtained from Fourier transformation of the proton free-induction decay (FID) signals.   
The proton nuclear spin-lattice relaxation curves $p(t)\equiv 1-F(t)/F(\infty)$ [recovery curves of the integrated FID signal $F(t)$] were measured by using an inversion recovery technique as a function of time $t$ ($>$ 5 ms) following an inversion pulse.
The equilibrium proton magnetization $F(\infty)$ is the integrated FID signal without an inversion pulse. 

\section{Experimental Results and Discussion}
\subsection{Development  of narrow FT NMR spectra}
 
\begin{figure}[t]
 \begin{center} 
\includegraphics[width=0.77\linewidth]{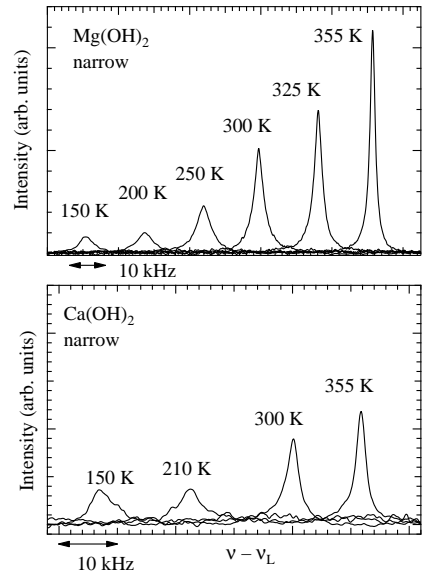}
 \end{center}
 \caption{\label{f2}
Temperature dependences of narrow NMR spectra at $\nu_\mathrm{L}$ = 42.5772 MHz for Mg(OH)$_2$ and Ca(OH)$_2$. 
Note that the spectra are displaced along the horizontal axis for clarity and that the frequency scale of the upper panel is different from that of the lower panel. The intensity of either narrow component increases on heating in a manner not satisfying Curie's law. 
 }
\end{figure}

Figure~\ref{f2} shows temperature dependences of the narrow NMR spectra at $\nu_\mathrm{L}$ = 42.5772 MHz for Mg(OH)$_2$ and Ca(OH)$_2$. 
Note that the spectra are displaced along the horizontal axis for clarity and that the frequency scale of the upper panel is different from that of the lower panel. 
The intensity of either narrow component increases with increasing temperature in a manner not satisfying Curie's law. 

\subsubsection{Mg(OH)$_2$}
   
\begin{figure}[t]
 \begin{center}
 \includegraphics[width=0.8\linewidth]{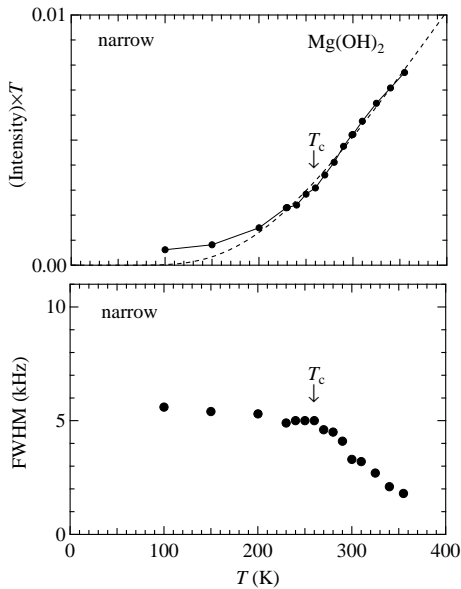}
 \end{center}
 \caption{\label{f3}
Mg(OH)$_2$: Temperature dependences of the product of the integrated intensity and temperature (upper panel) and the FWHM (lower panel) of the narrow component at  $\nu_\mathrm{L}$ = 42.5772 MHz. 
The intensity is normalized by the total intensity of the broad and narrow components at 355 K. 
The narrow component develops above $T_c$ = 260 K and is finite below $T_c$. 
The solid curve is a visual guide. 
The dashed curve indicates the result from least-squares fitting using a thermal-activation-type function $I_0$exp($-\Delta/k_\mathrm{B}T$) with $\Delta$ = 0.082 eV. 
 }
 \end{figure}

Figure~\ref{f3} shows temperature dependences of the product of the integrated intensity and temperature (upper panel) and the full width at half maximum (FWHM) (lower panel) of the narrow component for Mg(OH)$_2$. 
The intensity is normalized by the total intensity of the broad and narrow components at 355 K. 
The narrow component develops above about $T_c$ = 260 K, which is defined as the inflection point at which the solid curve changes from being concave downward to concave upward, in a manner not satisfying Curie's law. 
The temperature dependence of the narrow NMR spectrum was reversible in the temperature range of 100$-$355 K. 
No hysteresis with temperature was observed within the experimental accuracy.
The intensity of the narrow component was fit to a thermal-activation-type function $I$($T$) = $I_0$exp($-\Delta/k_\mathrm{B}T$) ($I_0$ and $\Delta$ are fitting parameters) with $\Delta$ = 0.082 eV, indicated by the dashed curve.  
The inflection point of $T_c$ = 260 K is inconsistent with that of $I$($T$) with $\Delta$ = 0.082 eV.
The activation function may not be the best fitting function.

Since Mg(OH)$_2$ is a proton conductor above room temperature~\cite{Freund},   
the narrow component may be assigned to mobile protons facilitated by lattice defects and activated by thermal energy. 
The activation energy of $\Delta$ = 0.082 eV in the narrow NMR spectrum in Fig.~\ref{f3} is 
about 24 times smaller than the activation energy of 2.0 eV for the proton conductivity~\cite{Freund} and 
about five times smaller than the excitation energy of 0.4 eV for a proton jump in a local Morse potential~\cite{FreundB}. 
The discrepancy between the estimated values suggests that 
the mechanism of the proton conduction below 355 K detected by the NMR is different from that of the proton transport above room temperature. 
Quantum tunneling of protons is a possible candidate for the proton conduction below room temperature.  

Figure~\ref{f3} (lower panel) shows the temperature dependence of the FWHM of the narrow NMR spectrum for Mg(OH)$_2$. 
Above $T_c$, the FWHM is about 3.3 kHz at 300 K. 
Below $T_c$, the FWHM is about 5.5 kHz at 100 K. 
Below $T_c$, the intensity is small but finite, and the FWHM is broadened but still be narrow.    
Thus, $T_c$ characterizes not a phase transition but a crossover phenomenon. 
 
\begin{figure}[h]
 \begin{center}
 \includegraphics[width=0.8\linewidth]{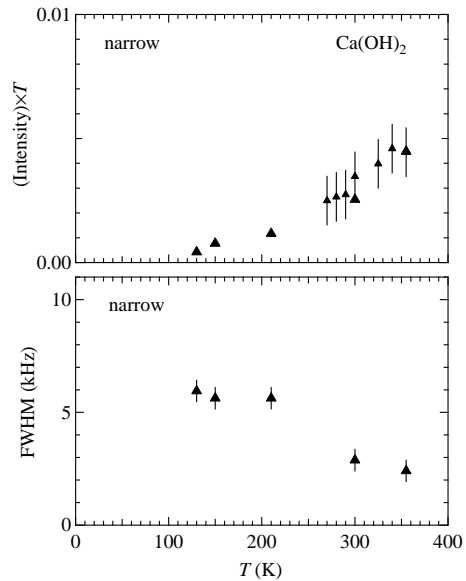}
 \end{center}
 \caption{\label{f4}
Ca(OH)$_2$: Temperature dependences of the product of the integrated intensity and temperature (upper panel) and the FWHM (lower panel) of the narrow component at  $\nu_\mathrm{L}$ = 42.5772 MHz.   
The intensity of the narrow component is less than 1$\%$ of the total intensity. 
 }
 \end{figure}        

\subsubsection{Ca(OH)$_2$}

Figure~\ref{f4} shows temperature dependences of the product of the integrated intensity and temperature (upper panel) and the FWHM (lower panel) of the narrow component for Ca(OH)$_2$. 
The temperature dependences of the intensity and the FWHM for Ca(OH)$_2$ are nearly the same as those for Mg(OH)$_2$. 
The ratio of the narrow component to the broad component in Ca(OH)$_2$ is less than that in Mg(OH)$_2$ at 355 K.  
Thus, the number of proton defects in Ca(OH)$_2$ may be less than that in Mg(OH)$_2$.
The difference in the defect density may be related to that in the proton conductivity. 
 
 % - - - - - - - - - - - - - - - - - - - - -
\subsection{Proton spin-lattice relaxation of narrow component}
\subsubsection{Mg(OH)$_2$}
  
For Mg(OH)$_2$, the recovery curves of the long tail of the FID signal $F(t)$ (the integrated intensity after 60 $\mu$s following an excitation pulse) were recorded 
as a function of the time $t$ following the inversion pulse in order to selectively observe the narrow NMR component.  
Although the theoretical recovery curve for $^1$H NMR should be a single-exponential function in a uniform system,  
all the recovery curves were of a nonexponential type, which is consistent with a previous report~\cite{pNMR}.  
We analyzed the recovery curves using a stretched exponential function with a variable exponent $\beta$,
\begin{equation}
p(t)=p(0)\text{exp}\Bigl[-\Bigl({t\over {\tau_1}}\Bigr)^\beta \Bigr],
%p(t)=p(0){e}^{-(t/\tau_1)^\beta},  
\label{eq.1}
\end{equation}
where $p$(0), $\tau_1$, and $\beta$ are fitting parameters.  

  \begin{figure}[h]
 \begin{center}
 \includegraphics[width=0.75\linewidth]{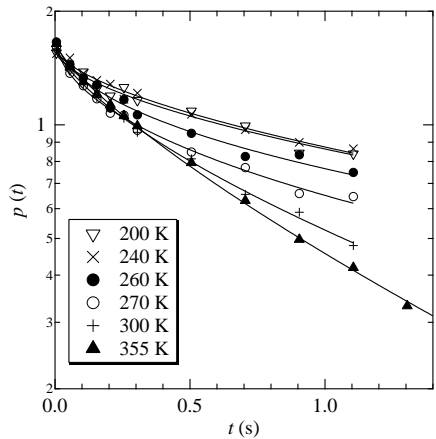}
 \end{center}
 \caption{\label{f5}
Mg(OH)$_2$: Proton spin-lattice relaxation curves of the narrow component at 42.5772 MHz. 
Solid curves are the least-squares fitting results using the stretched exponential function with a variable exponent given by Eq.~(\ref{eq.1}).  
}
 \end{figure}  
 
 \begin{figure}[h]
 \begin{center}
 \includegraphics[width=0.8\linewidth]{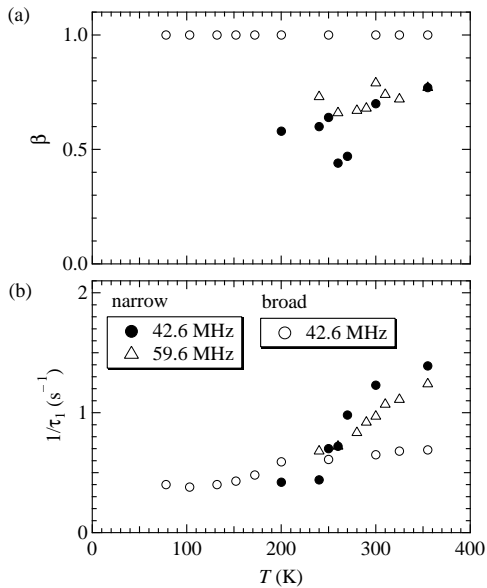}
 \end{center}
 \caption{\label{f6}
Mg(OH)$_2$: Temperature dependences of the stretched exponent $\beta$ (a) and the proton spin-lattice relaxation rate 1/$\tau_1$ (b) of the narrow component at 42.5772 and 59.6 MHz. For comparison, the stretched exponent $\beta$ (a) and the proton spin-lattice relaxation rate 1/$T_1$ (b) of the broad component are also reproduced from a previous report~\cite{Itoh}.    
 }
 \end{figure}      

Figure~\ref{f5} shows the temperature dependence of the proton spin-lattice relaxation curve of the narrow component. 
The sold curves are the least-squares fitting results using Eq.~(\ref{eq.1}).  

Figure~\ref{f6} shows the temperature dependences of the variable exponent $\beta$ (a) and 1/$\tau_{1}$ (b). 
A value of $\beta <$ 1 indicates an inhomogeneous relaxation process in space in the temperature range of 200$-$355 K. 
Part of the protons quickly relax in $t/\tau_1 <$ 1. 

The proton spin-lattice relaxation rate 1/$\tau_1$ monotonically decreases on cooling. 
Above about 260 K, the stretched exponential relaxation rate 1/$\tau_1$ of the narrow component is larger than the single-exponential relaxation rate 1/$T_1$ of the broad component~\cite{Itoh}.
The temperature dependence of 1/$\tau_1$ is different from that expected from the slowing-down effects on the spin-spin correlation time $\tau_c$ 
in the Bloembergen$-$Purcell$-$Pound (BPP) model~\cite{BPP}.    
The frequency dependence of the fluctuation spectrum in the BPP model is of a Lorentzian type.
The relaxation rate due to the BPP fluctuations should show a peak at a temperature $T^*$ satisfying $\tau_c$($T^*$) = 1/2$\pi$$\nu_\mathrm{L}$.  

\subsubsection{Ca(OH)$_2$}
  
\begin{figure}[h]
 \begin{center}
 \includegraphics[width=0.75\linewidth]{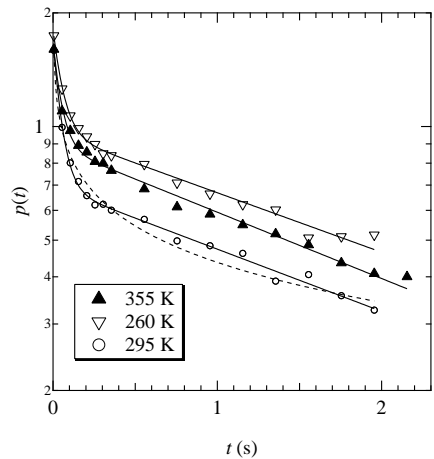}
 \end{center}
 \caption{\label{f7}
Ca(OH)$_2$: Proton spin-lattice relaxation curves of the narrow component at 42.5772 MHz. 
Solid curves are the least-squares fitting results using the double-exponential function given by Eq.~(\ref{eq.2}).  
The dashed curve is the least-squares fitting result using the stretched exponential function given by Eq.~(\ref{eq.1}).  
The double-exponential function reproduces the abrupt change in the relaxation curve better than the stretched exponential function. 
}
 \end{figure}

For Ca(OH)$_2$, all the recovery curves were of a nonexponential type, which is consistent with a previous report~\cite{T1Ca}.  
The iterative fitting using Eq.~(\ref{eq.1}) led to a small exponent $\beta$ $<$ 0.5 and the convergence was poor. 
Thus, the stretched exponential function given by Eq.~(\ref{eq.1}) could not accurately reproduce the recovery curves,
as shown by the dashed curve in Fig.~\ref{f7}. 
We obtained better fitting results by using a double-exponential function,
\begin{equation} 
p(t)=p_L(0)\text{exp}\Bigl(-{t\over {T_{1L}}}\Bigr)+p_S(0)\text{exp}\Bigl(-{t\over {T_{1S}}}\Bigr), 
\label{eq.2}
\end{equation}
where $p_L$(0), $T_{1L}$, $p_S$(0), and $T_{1S}$ are fitting parameters.  

 \begin{figure}[h]
 \begin{center}
 \includegraphics[width=0.8\linewidth]{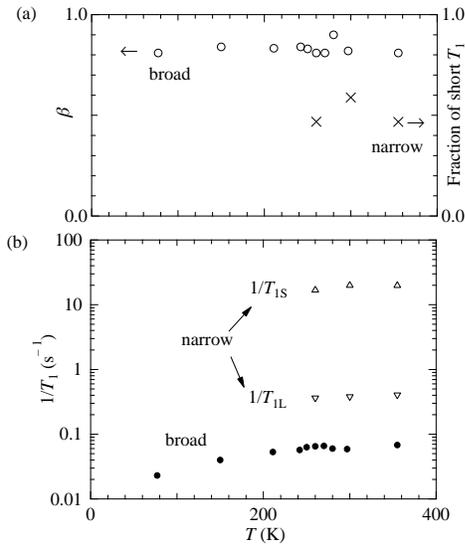}
 \end{center}
 \caption{\label{f8}
Ca(OH)$_2$: Temperature dependences of the fraction of the short relaxation component $p_S$(0)/[$p_S$(0)+$p_L$(0)] (a) and the proton spin-lattice relaxation rates 1/$T_{1L}$ and 1/$T_{1S}$ (b) of the narrow component at 42.5772 MHz. For comparison, the temperature dependences of the stretched exponent $\beta$ (a) and the proton spin-lattice relaxation rate 1/$T_1$ (b) of the broad component are also shown.    
 }
 \end{figure}      

Figure~\ref{f7} shows the temperature dependence of the proton spin-lattice relaxation curve of the narrow component at 42.5772 MHz. 
The solid curves are the least-squares fitting results using the double-exponential function given by Eq.~(\ref{eq.2}). 
The dashed curve is the least-squares fitting result using the stretched exponential function given by Eq.~(\ref{eq.1}).  
The fitting results obtained using the double-exponential function are better than those obtained using the stretched exponential function.

Figure~\ref{f8} shows the temperature dependences of the fraction of the short relaxation component $p_S$(0)/[$p_S$(0)+$p_L$(0)] (a) and the proton spin-lattice relaxation rates 1/$T_{1L}$ and 1/$T_{1S}$ (b) of the narrow component. 
For comparison, we measured the proton spin-lattice relaxation rate 1/$T_1$ of the broad component for Ca(OH)$_2$ in a similar manner to that for Mg(OH)$_2$~\cite{Itoh}, which is also shown in Fig.~\ref{f8}.  

Both 1/$T_{1L}$ and 1/$T_{1S}$ for the narrow component show weak dependence on the temperature
and are larger than the relaxation rate 1/$T_1$ of the broad component.
No slowing-down effect was found in any of the nuclear spin-lattice relaxation components.

The large difference in the proton spin-lattice relaxation between the narrow and broad components is a characteristic of Ca(OH)$_2$.  
The proton fluctuation spectrum of the narrow component for Ca(OH)$_2$ may have a different nature from that for Mg(OH)$_2$.    
    
\subsection{NMR spectra with variable Larmor frequency for Mg(OH)$_2$}   
 
\begin{figure}[t]
 \begin{center}
 \includegraphics[width=0.85\linewidth]{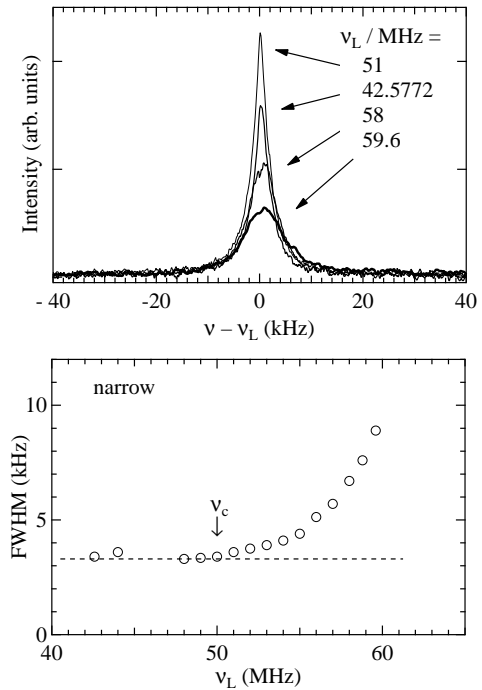}
 \end{center}
 \caption{\label{f9}
Mg(OH)$_2$: Larmor frequency dependences of the narrow NMR spectrum (upper panel) and the FWHM (lower panel) at 42.5772$-$59.6 MHz and 300 K. 
The FWHM rapidly increases above about $\nu_c$ = 50 MHz.   
 }
 \end{figure}

Figure~\ref{f9} shows the Larmor frequency dependences of the narrow component (upper panel) and the FWHM (lower panel) for Mg(OH)$_2$ over the frequency range of 42.5772$-$59.6 MHz at 300 K. 
The linewidth of the narrow component rapidly increases above about $\nu_c$ = 50 MHz, which we identify as the characteristic frequency of the fluctuation spectrum of the extrinsic protons.  
The frequency dependence was reversible and exhibited no hysteresis within the experimental accuracy. 
The temperature of about 300 K is sufficiently high for the magnetic field effect on the nuclear magnetism to be negligibly small. 
Figure~\ref{f9} shows the pure frequency dependence of the NMR spectrum rather than the magnetic field effect. 
Measurement at a higher Larmor frequency provides a higher-speed snapshot of fluctuating protons. 
The snapshot taken above $\nu_c$ = 50 MHz indicates a widening narrow spectrum while the narrowing effect is still active. 
The temperature dependences of the intensity of the narrow and broad components at 59.6 MHz were similar to those at 42.5772 MHz in Fig.~\ref{f9}.      
    
 \subsection{Temperature-induced $^{1}$H NMR signals}  
 
 The temperature-induced $^{1}$H NMR spectra and the two-component spectra have been observed for MgH$_2$~\cite{Corey1}, LiBH$_4$~\cite{Corey2}, NaMgH$_3$~\cite{Corey3}, NaH~\cite{Corey4}, and the KH$_2$PO$_4$ family~\cite{TDA, KDP}.   
 
% ------------ 
A pulsed proton NMR study of ball-milled magnesium hydrate (MgH$_2$) with Nb$_2$O$_5$ additive produced
narrow and broad NMR spectra above room temperature~\cite{Corey1}. 
MgH$_2$ is a promising hydrogen storage material. 
The narrow component was assigned to mobile protons~\cite{Corey1}.
Subsequent studies by transmission electron microscopy (TEM) support the assignment of the NMR spectra. 
The TEM imaging suggested that the protons diffuse at the interface between MgH$_2$ and Nb$_2$O$_5$ grains~\cite{Isobe}. 

For Mg(OH)$_2$ and Ca(OH)$_2$, part of the protons may depart from the XGT sites and may be itinerant among the defect sites and/or the intersites above $T_{c}$. 
The number of itinerant protons may increase with increasing temperature above $T_{c}$. 
The increase in the intensity of the narrow NMR spectrum is similar to that in the number of itinerant protons. 
The extrinsic protons may be such itinerant protons. 
Irreversible surface degradation of the powder grains is excluded because the two-component NMR spectra were reversible in the heat cycle.  

The characteristic frequency $\nu_c$ of 50 MHz has not been reported so far for the proton conductors.  
The Larmor frequency dependence in Fig.~\ref{f9} may be a key to understanding the temperature-induced $^{1}$H NMR signals
and the spin fluctuation spectrum of the extrinsic protons.  

%------------------------
\section{Conclusions}

We observed the temperature-induced narrow component of the proton NMR spectrum for Mg(OH)$_2$ and Ca(OH)$_2$ powders. 
Above the characteristic temperature $T_c$ of 260 K, the NMR line develops while narrowing. 
Above the characteristic frequency $\nu_c$ of 50 MHz, the NMR line decreases while broadening. 
No slowing-down effect was observed in any of the components of the proton spin-lattice relaxation. 
The fluctuation spectrum of the extrinsic protons is different from that obtained from the conventional BPP model.

\end{document}